\documentstyle[12pt]{article}
\def\beq{\begin{equation}}
\def\eeq{\end{equation}}

\newcommand{\ev}[1]{\langle #1 \rangle}
\newfont\figfont{cmr7 scaled 1200}
\begin{document}
\title{The ACCMM Model and the Heavy Quark Expansion
\thanks{This work is supported
in part by funds provided by the U.S.
Department of Energy (DOE) under contract \#DE-AC02-76ER03069 and in
part by the Texas National Research Laboratory Commission
under grant \#RGFY92C6.\hfill\break
\vskip 0.05cm
\noindent $^{\dagger}$National
Science Foundation Young Investigator Award.\hfill\break
Alfred P.~Sloan
Foundation Research Fellowship.\hfill\break
Department of Energy Outstanding Junior
Investigator Award.\hfill\break
CTP\#2262, hep-ph/9312257\hfill December 1993}}
\renewcommand{\baselinestretch}{1.0}
\author{Csaba Cs\'aki and Lisa Randall$^{\dagger}$ \\
Massachusetts Institute of Technology\\
Cambridge, MA 02139\\
}
\date{}
\maketitle
\vskip-5in
December1993 \hfill  MIT-CTP\#2262
\vskip5in
\renewcommand{\baselinestretch}{1.2}
\abstract{
The ACCMM model predicts the lepton spectrum from $B$ meson decay
by assuming the meson disintegrates into a spectator quark of definite
mass and momentum distribution and an off shell $b$ quark whose
decay leptons (boosted into the rest frame of the meson) determine
the lepton spectrum.  In this letter, we show that one can define
a model dependent $b$ quark mass so that the spectrum derived from the ACCMM
model agrees very well with the free quark decay spectrum far from the
endpoint.
Near
the  endpoint,  there is some disagreement, indicating the result is more model
dependent.
The integrated spectra are   however very nearly identical.  These results
are in accordance with expectations based on the heavy quark effective theory.
 We conclude that for LEP experiments, free quark
decay might be as general and more simple than the ACCMM ansatz for modeling
the
inclusive charged lepton spectrum from  $B$ meson decay.
\thispagestyle{empty}
\newpage

\section{Introduction}
\def\lom{$\Lambda/m$\ }
Much attention has been devoted recently to the study of the lepton spectrum
from inclusive semileptonic heavy hadron decay
  \cite{chay,shif,mw,flukes}. The key idea is that when  a
heavy hadron decays, the operator product expansion  (OPE) and the heavy quark
effective theory (HQET)
 may be employed  to derive the decay spectrum. These methods were employed
in the study of  inclusive  semileptonic
decay $H_b \to X l \overline{\nu}_L$ where $H_b$ is a bottom hadron.
The result (away
from the endpoint of the spectrum) is
that the inclusive differential decay  rate ${d \Gamma \over dE}$
may be  expanded in $\Lambda/m$, where $\Lambda$ is a QCD related
scale of order 500 MeV, and $m$ is the mass of the heavy quark. The leading
term (zeroth order in $\Lambda/m$) is the free quark decay spectrum, the
subleading term vanishes, and the subsubleading term involves parameters
from the heavy quark theory, but should be rather small  for $b$ quark decay,
as it is  of order $(\Lambda/m)^2$.

However, it is not always apparent how models reproduce this result. For
example,
it is not manifest in the model of Altarelli, Cabbibo, Corb\`o,
Maiani \cite{accmm}, and Martinelli, hereafter referred to as ACCMM,
in which  the differential decay spectrum of the charged lepton  for inclusive
semileptonic  $B$ meson decay is derived under some assumptions
about the bound state. In this paper, we show
that the linear terms in the differential distribution in the
region far from the endpoint  or of  a suitably averaged spectrum  do vanish
(contrary
to the claim in  ref. \cite{shif}).   We present
a detailed discussion of this model and  show that by a suitable {\it model
dependent}
definition of the $b$ quark mass, the linear terms in \lom may be eliminated so
that the predictions of the  ACCMM model agree well with  those of the heavy
quark effective theory (HQET).  We  first do the calculation directly from the
ACCMM model, without
invoking the formalism of HQET or the OPE. By exploring this particular model,
we
hope to clarify the relation between models and the HQET result.  Many of these
ideas  were discussed independently in ref. \cite{mw}.

Having shown that free quark decay reproduces the ACCMM spectrum up
to small corrections (of order $(\Lambda/m)^2$ which should be of  the order of
a percent),
we discuss the question of whether simple  free quark decay (with
the $b$ quark mass the free parameter) is as viable
for modeling the lepton spectrum as the ACCMM model.
We will see that the two models agree well far from the endpoint but deviate
within a
region of approximately $2p_f$ from the endpoint.
We consider the ACCMM spectrum
in detail and also compare the results of an averaged ACCMM model to an
averaged free quark model.   We ask the question
of whether smearing is required to get good agreement
between the models, and if so, how much.  It is useful to understand
the endpoint region in a model in which there are no singularities introduced
by using
the OPE in the region where it does not converge. The examination of
a particular model provides  a complementary approach to determining
the necessary averaging which might be required in the endpoint region.
Our conclusion is that free quark decay probably works as well  for experiments
at LEP, for example,  as the ACCMM model   and in fact
has the same number of independent parameters (if the spectator mass of the
ACCMM model
is
taken to be fixed). Whether the details of the endpoint region
are important  will however depend
on the  the  application of the models and the accuracy which is required.

Although we are considering free quark decay, there do exist  higher order
corrections.  If accuracy
better than a percent is required these higher order corrections
 must be included. These were considered
in detail in refs. \cite{shif,mw}. The focus of this paper is not these small
corrections,
but to show that the linear correction terms can be made to  vanish, and that
free quark
decay models fairly well the ACCMM spectrum (so long as the detailed structure
of the endpoint is not important). However, we briefly comment in the
conclusion on the generality of the higher order corrections of the ACCMM
model.
We also neglect perturbative QCD corrections in our discussion, but these
can be readily incorporated \cite{accmm,qcd}. Finally, we will consider a decay
to the massive $c$ quark,
but our conclusions hold for a massless final state quark as well. The shape of
the
spectrum obtained in the massless case would however be significantly modified
by perturbative QCD.

We note that  the ACCMM model was
developed to consider in detail the endpoint of the lepton spectrum.
 As we will
discuss, the ACCMM model  is simply not presented in a form where its relation
to the HQET
result is obvious.  In fact, it was explicitly constructed to avoid mention of
a $b$ quark mass.
The LEP
experiments appropriate this model, which was designed to study
the endpoint, in order to estimate a systematic error in modeling  the full
spectrum.
Our conclusion is that far from the endpoint, free quark decay is as good
as the ACCMM model. The only model dependence comes in the treatment of the
endpoint region. Any result  depending on the details of this region is of
course model dependent and unreliable. However, if it is only the distribution
smeared over energy which is relevant, we will see that the ACCMM model
and free quark decay give the same predictions at about the percent level.

The outline of this paper is as follows. In the first section, we show how the
ACCMM
spectrum can be analytically expanded in $b$ quark mass when far from the
endpoint region. In
the second section, we review the relevant results from the HQET and see
how they apply to the ACCMM model. Section 4 considers in more detail the
endpoint region of the spectrum, as well as the total integrated spectrum.   We
conclude in the final
section.

\section{The ACCMM Model}
In order to incorporate the fact that when a $b$ quark decays its spectrum
is not simply that from free quark decay, because the quark is in a bound
state,
 the ACCMM model treats the decay of the
$B$ meson as a disintegration into a spectator quark of given mass and
 distribution of momentum plus a $b$ quark . The decay spectrum
is determined by the kinematical constraints on the $b$ quark. This should
incorporate
at least some of the corrections related to the fact that the $b$ quark
which decays is not free, but in a bound state.

As we will discuss further in the following section, the heavy quark theory
predicts that any model in accordance with its assumptions will give
rise to a decay spectrum which up to small corrections (of order
$(\Lambda/m)^2$)
agrees with the free quark decay spectra, at least when sufficiently far from
the endpoint (or when a suitable averaging procedure is applied). Since
the ACCMM model does not obviously violate any of the underlying assumptions of
a
heavy quark model,  we should expect the resulting decay spectrum to
very nearly agree with  that of the decay of a free quark when far from the
endpoint.
Nonetheless, because the model does not incorporate the underlying QCD
invariance,
the predictions from this model are not compatible with the OPE/HQET result
at the $(\Lambda/m_b)^2$ level.

 In ref. \cite{accmm}, the results were obtained  numerically,
so it is difficult to identify the heavy quark mass dependence. Here,
 we do the heavy quark mass expansion
explicitly. We first do the calculation in terms of the invariant mass
of the $b$ quark inside the $B$ meson.  We reexpress the answer in
terms of the parameters
 used
in the paper, namely $m_B$, the $B$ meson mass and $p_f$, a parameter
characterizing their assumed form for nonperturbative corrections.
However, in this form, one cannot say anything about heavy quark
mass corrections because we have not even defined the $b$ quark mass.
We show that if we define $m_b$ to be the average value
of the energy of the $b$ quark,  the linear terms in the expansion
of the differential distribution  in powers of $\Lambda/m$ do vanish. It is
straightforward to generalize
this result to arbitrary spectator and final state quark masses.

\subsection{The Model}

\def\beq{\begin{equation}}
\def\eeq{\end{equation}}
The ACCMM model assumes a spectator quark of fixed mass $m_{sp}$ with
a momentum distribution given by $\phi(|p|)$. For simplicity, this is
taken to be  Gaussian:

\beq
\phi(|p|)={4 \over \sqrt{\pi} p_f^3}\exp\left(-{|p|^2\over p_f^2}\right)
\eeq
which is normalized so that the integral over all momenta
of $\phi(|p|) p^2$  (not only those which
are allowed kinematically) is 1.

Assuming conservation of energy and  momentum tells us that
when the momentum of the spectator quark has
magnitude $|p|$ and mass $m_{sp}$,  the energy of the
heavy quark will be
\beq
E_W=m_B-\sqrt{p^2+m_{sp}^2}
\eeq
where $m_B$ is the $B$ meson mass.

The invariant mass of the $b$ quark  will then be
\beq
W^2=m_B^2+m_{sp}^2-2m_B\sqrt{p^2+m_{sp}^2}
\eeq

According to these assumptions, the distribution of lepton energy can
be determined by boosting back the decay products of the $b$ quarks of
invariant
mass $W$ (in the $b$ quark rest frame) to the rest frame of the meson and
averaging
over momenta.

We define
\begin{eqnarray}
x&=&{2 E_e \over m_b}\\
\epsilon&=&{m_f^2 \over m_b^2}\\
x_m&=& 1-\epsilon
\end{eqnarray}

Here, $m_f$ is the final state quark mass.
Then the lepton spectrum from the decaying $b$ quark in its rest frame is
\beq
{d \Gamma(m_b,E) \over d E}={G_F^2 m_b^4 \over 48\pi^3}{x^2 (x_m-x)^2 \over
(1-x)^3} \left[(1-x)(3-2x)+(1-x_m)(3-x)\right]
\eeq

Define $\gamma=E_W/W$, $\beta=p/E_W$.
The  spectrum of leptons in the ACCMM model is determined from
\begin{eqnarray}
{d \Gamma_B \over dE}&=&\int_0^{p_{max}} dp p^2 \phi(|p|) \int {1\over \gamma}
{d^2 \Gamma(W,E') \over dE' d \cos\theta}
dE'  \times\\
&& \ \ \ d\cos\theta \int{d \cos \theta_p \over 2} \delta(E-\gamma E'-\gamma
\beta E' \cos \theta_p)
\end{eqnarray}
Here we have  assumed the lepton is massless and we have defined the angle
$\theta_p$
(associated with the distribution in momentum) with
respect to the angle of the decaying lepton, for each of the orientations.  It
is straightforward to do the $\theta_p$ integral to derive
\beq
{d \Gamma_B \over dE}=\int dp p^2 \phi(|p|)
{1 \over 2 \beta \gamma ^2} \int {d^2 \Gamma(W,E') \over dE' d\cos \theta} d
\cos \theta{dE' \over E'}
\eeq
We use the fact that $\beta \gamma^2=p E_W/W^2(p)$ and integrate
over $\cos\theta$ to obtain
\beq
{d \Gamma_B \over dE}=\int dp p^2 \phi(|p|){W^2 \over 2p E_W}
\int_{E_-}^{E_{max}} {dE' \over E'} {d \Gamma(W,E') \over dE'}
\eeq
where
\begin{eqnarray}
E_{\mp}&=&{E W \over E_W\pm p}\\
E_{max}&=&\frac{m_B-m_{sp}}{2}\left (1-\frac{m_f^2}{(m_B-m_{sp})^2}\right)\\
p_{max}&=&{m_B \over 2}-{m_f^2 \over  2m_B -4E }
\end{eqnarray}
where the former are determined by the $\delta$ function and the fact that
the decay product comes from a quark of ``mass" $W$ and the latter comes
from requiring that $E_{max}>E_-$ and is given for $m_{sp}=0$.
This is the result of ACCMM (up to the $\gamma$ factor, which is a small
correction).

\subsection{Heavy Quark Expansion Applied to the Model}
We now consider the implications of the expression we have just derived. We
first
 restrict our attention to massless final state and massless  quarks.
Furthermore, we consider the distribution far
 from the endpoint  and assume small $p_f$ (so that $E_{max}=E_+$) . We then
have
\beq
{d \Gamma_B \over dE}=\int_0^{p_{max}} \phi(|p|) p^2 {W^2 \over 2p E_W} {G_F^2
W^4 \over 48 \pi^3}
\int_{E_-}^{E_+}{dE' \over E'}\left({2E \over W}\right)^2 \left(3-{4E' \over
W}\right)
\eeq

We can explicitly evaluate the $E'$ integral, to get
\beq\label{wfcn}
{d \Gamma_B \over dE}={G_F^2 E^2 \over 24 \pi^3}\int_0^{p_{max}}{dp \over
E_W}\left(
6 E_W W^2-8 E E_W^2-{8 \over 3} p^2 E \right) p^2 \phi(|p|)
\eeq
where it should be borne in mind that both $W$ and $E_W$ are functions of $p$.

We now consider this expression when $p_f$ is small.  The value of $p_{max}$
 can be taken to be approximately $\infty$ so far as the integral goes, making
a negligible error. We also see that only $p\ll m_B$ will contribute
significantly
to the integral (in fact less than  about 2$p_f$), so that we can also expand
the denominator in $p/m_B$.  We expand in $p$ and do the integrals,
obtaining
\beq
{d \Gamma_B \over dE}={d \Gamma_B^0 \over dE}+{G_F^2 E^2 m_B^2 \over 12 \pi^3
\sqrt{\pi}}\left(8{ E \over m_B}-12\right){p_f \over m_B}+O\left({p_f^2\over
m_B^2}\right) {d \Gamma_q \over dE}
\eeq
where the first term represents the decay of a quark of mass $m_B$.

So if we expand in terms of the meson mass, it looks
like the differential decay distribution is that for a free quark of mass $m_B$
which decays plus a linear correction term of order $p_f/m_B$. However,
it is clear that in this form we learn nothing about the heavy quark expansion
since the answer is expressed in terms of the {\it meson} rather than the {\it
quark} mass.
In fact, at this point, it is not even clear what the quark mass means in this
model.
Nonetheless, we can {\it define} the mass of the $b$ quark to be that
mass for which this expression looks like free quark decay up to quadratic
corrections. This definition is not as random as it sounds, since from equation
(\ref{wfcn}) we see that the final spectrum is  a function of $W$ and $E_W$.
Therefore, the way to eliminate the linear terms in the decay distribution is
to define the quark mass so that neither $W$ nor $E_W$ has linear corrections;
that is define the quark mass $m_b$ by
\beq
m_b=\langle E_W(p) \rangle
\eeq
Up to quadratic terms in $p$, we have
\beq
\langle W \rangle =\langle E_W \rangle=m_B-\langle p \rangle\equiv m_b
\eeq
Because
\beq
\langle f(E_W) \rangle = f(m_B)-\langle p \rangle f'(m_B) +
O(p^2)=f(m_b)+O(p^2/m_b^2)
\eeq
we see that when we define  the $b$ quark mass this way that
any function $f(E_W)$  which can be expanded in $p_f/f(E_W)$ will
be equal to $f(m_b)+O(p_f^2/m_b^2)$. It is easy to check
that when we express the distribution as a function of $m_b$
there is no linear correction.
{}From this viewpoint the vanishing of the linear term is not very deep. It is
just
the statement that all the $p$ dependence is through $E_W$,  so by defining
$m_b$ to be
its average, we eliminate the linear corrections to the free quark differential
decay spectrum.
Notice this is true even though
the $b$ quark is not on mass shell.
 In fact,  we checked that if one  constructed a nonrelativistic quark model
with
the $b$ quark on shell and a Gaussian momentum distribution that  the
spectrum falls about halfway between the ACCMM prediction and a free quark
model.

This analysis makes clear why there is always a definition of $b$ quark
mass for which the linear correction terms vanish.   Up to
terms quadratic in $p^2/m_b^2$,  the $p$ dependence is all through
$E_W$.  Once we define the average of $E_W $ to be
the $b$ quark mass, all linear corrections are eliminated (so long
as we are in a regime where the expansion in momentum is legitimate).
Notice that in this model, the relation between the $b$ quark mass and the
quark mass defined at high energy can in principle differ by an amount of order
$\Lambda_{QCD}$.  However, this is an artifact of the model.

It is easy to see that this analysis can readily be extended to the case
when  the spectator quark mass is nonzero.
We then have
\beq
\langle E_W \rangle=m_B-\langle \sqrt{m_{sp}^2+p^2}\rangle
\eeq
Again, because everything can be expressed as a function of $W$ (up to small
corrections of order $(\Lambda/m)^2$, there
is only a single quantity we need to know. This is true despite the
fact that  naively, it appears that $m_{sp}$ and $p_f$ are independent
parameters.
We see that again, everything can be expressed in terms of $\ev{E_W}$. Of
course, at higher
order in $(\Lambda/m)^2$, this is no longer the case. This is briefly discussed
in the conclusion.

To extend the analysis to the case of nonzero final state quark mass, one
can proceed as above and explicitly do the integral over $E'$. However, it
is simpler to proceed as follows. We observe that
\beq
E_+-E_-={2 E p \over W}
\eeq
which is small if $p$ is small. Assuming that the integrand is a smooth
function
(ie we are far from the region near the endpoint where it varies rapidly)
we have
\begin{eqnarray}
{W^2 \over 2p E_W}\int_{E_-}^{E_+} {d \Gamma_q (W,E')\over dE'} {dE'\over
E'}&\approx&{W^2 \over 2p E_W}{d \Gamma_q(W, E) \over dE} Log\left({E_W+p \over
E_W-p}\right)\nonumber
\\ &&\approx
\left({1 \over \gamma}\right)^2{d \Gamma_q(W,p, E)\over dE}
\end{eqnarray}
The $p$ integral averages the differential distribution.
Because it only depends on $E_W$ (up to order $p^2/m^2$), the result is the
same as before;
linear corrections vanish if we choose $m_b=\ev{E_W}$.
Notice that in fact  that any choice which differed from this one
by terms of order $p^2/m_ b^2$ would also suffice.
(Here we chose  the velocity of the $b$ quark to be
 $v=(1,0,0,0)$.) The above argument is of course
very general. As long as the function is varying smoothly and
one is sufficiently far from the endpoint so that $E_{max}=E_+$, one  expects
corrections to the differential distribution to occur only at order
$(p_f/m_b)^2$.
Again we see that it is model indpendent, that is independent of the detailed
form
of the momentum distribution, $\phi(|p|)$.

Notice the same argument can be used to show that the ACCMM distribution
agrees exactly with that for a free quark
of mass $m_B-m_{sp}$ for $p_f=0$.   We therefore do not present results for
$p_f=0$ in what
follows.

We conclude that the differential distributions should agree well
between the ACCMM model and free quark decay with the $b$ quark
mass determined as above.  In Figs 1 and 2
we illustrate the agreement for $p_f=.15,.30, m_{sp}=.15,m_{final}=1.5$.
The discrepancy between the curves grows with $p_f$.
We choose these values for $p_f$ as they are the ones used in \cite{accmm}.

\begin{figure}
\let\picnaturalsize=N
\def\picsize{4.0 in}
\def\picfilename{pf015.ps}
\ifx\nopictures Y\else{\ifx\epsfloaded Y\else\input epsf \fi
\let\epsfloaded=Y
\centerline{\ifx\picnaturalsize N\epsfxsize \picsize\fi
\epsfbox{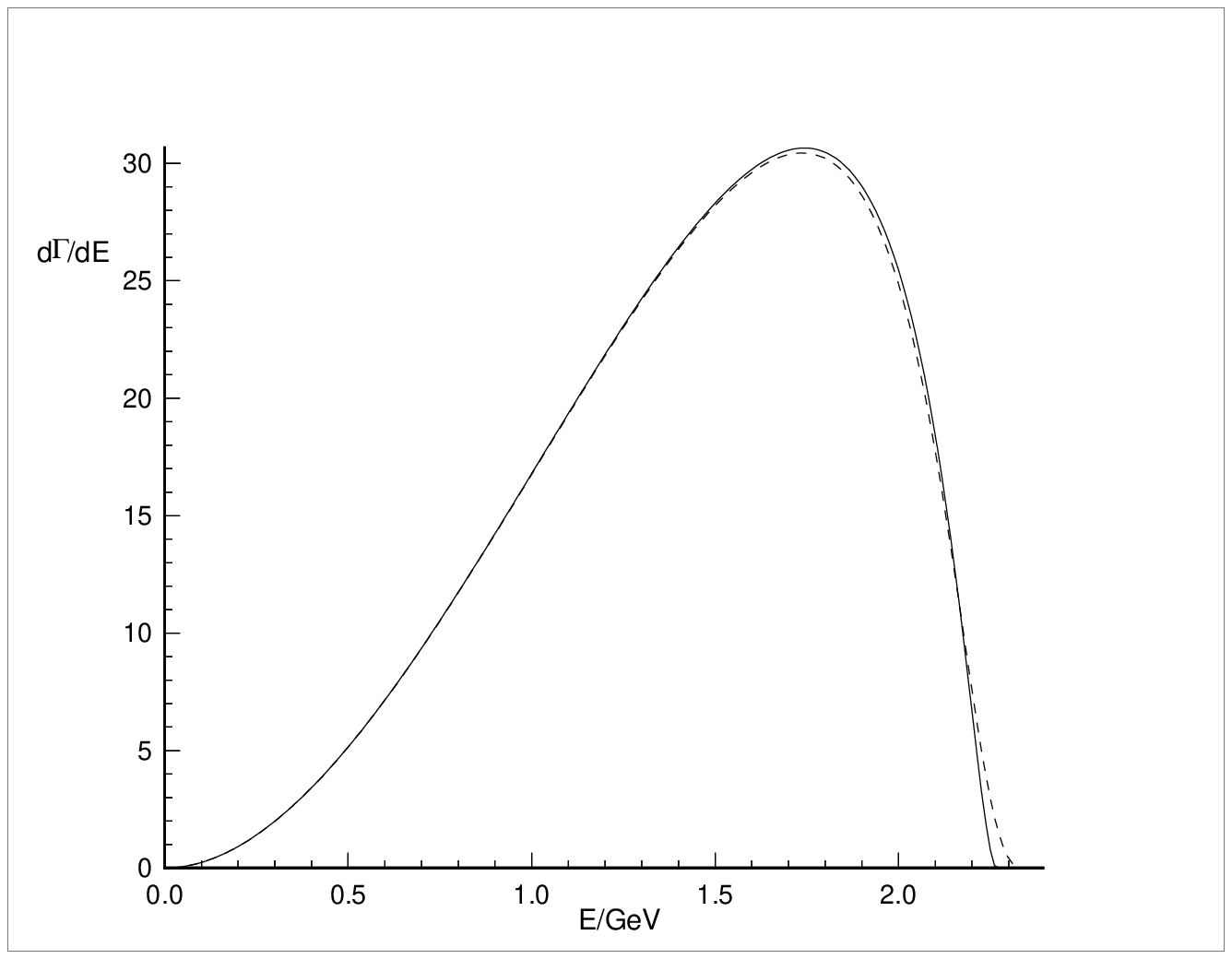}}}\fi

{\figfont  Figure 1: The inclusive differential semileptonic decay rate (in
units of $10^{-12}V_{cb}^2$) for the free quark model (solid curve) and the
ACCMM model (dashed curve). The parameters are $p_f=0.15,\; m_{sp}=0.15,\;
m_f=1.5$.}

\let\picnaturalsize=N
\def\picsize{4.0in}
\def\picfilename{pf03.ps}
\ifx\nopictures Y\else{\ifx\epsfloaded Y\else\input epsf \fi
\let\epsfloaded=Y
\centerline{\ifx\picnaturalsize N\epsfxsize \picsize\fi
\epsfbox{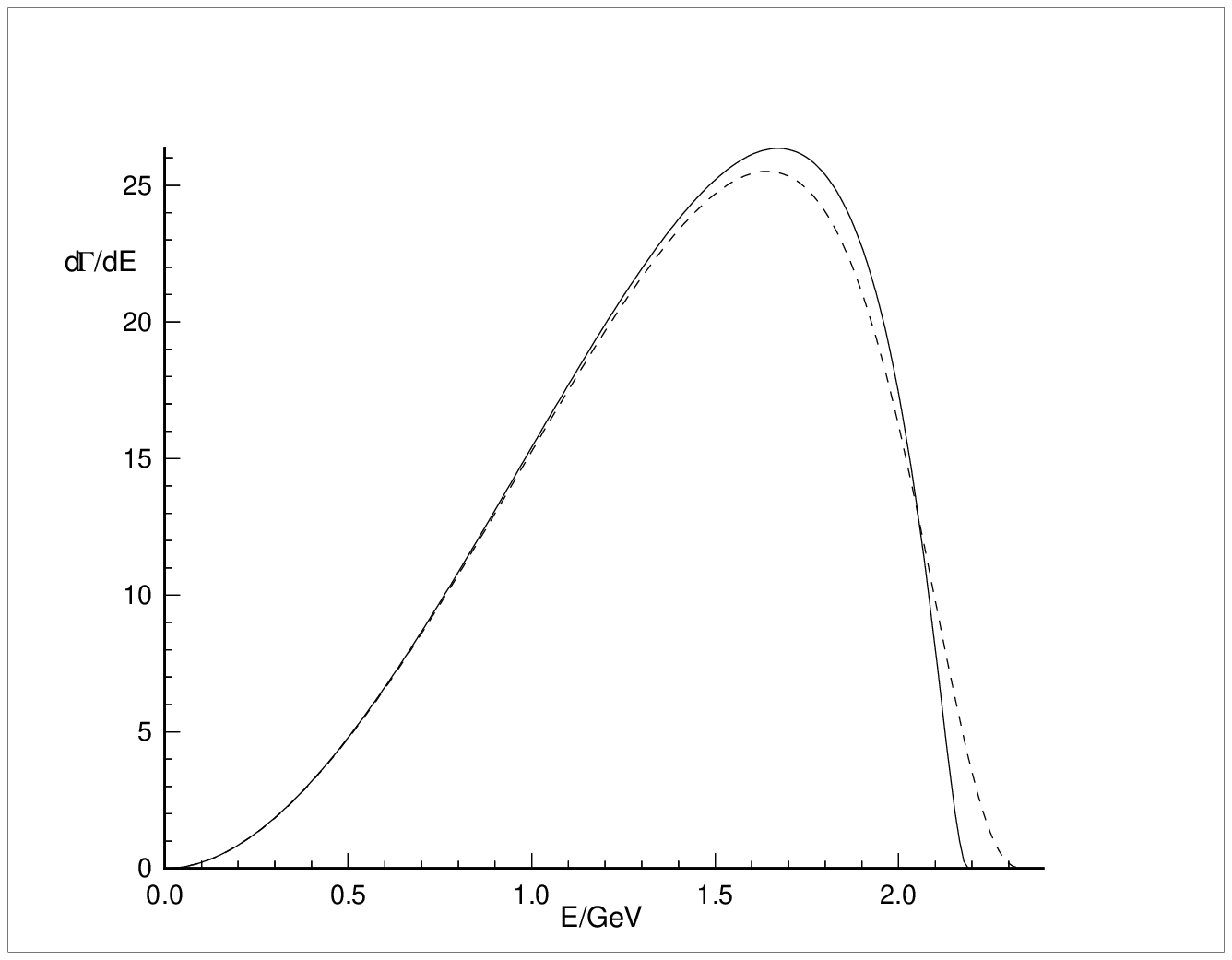}}}\fi

{\figfont  Figure 2: The same as in fig. 1  for $p_f=0.3$.}

\end{figure}

Notice we could also have considered a massless final state quark. We have
also checked our result for this case. Notice also that throughout this
paper we are neglecting perturbative QCD corrections. We did however check
that our integrated spectra agree with the numbers quoted in ACCMM when
the QCD corrections are incorporated. For a massless quark, the detailed
form of the QCD corrections needs to be incorporated. For a massive quark,
it is essentially an overall factor.

We will consider further the integrated spectra in Section 4. We first briefly
review
the HQET result.

\section{HQET Prediction for Spectrum}

The first paper to discuss the lepton spectrum in the context of the heavy
quark
theory is Chay {\it et al.} \cite{chay}. By using the operator product
expansion and
then the heavy quark effective theory, they showed how to  rigorously derive
both the leading order QCD result (free quark decay) and the correction
terms.   The correction terms are small (of order ($\Lambda/m)^2 $) and depend
on an unknown
parameter, so our primary interest will be the leading and subleading order
corrections.
  Most of this
section is review from ref. \cite{chay,shif,mw,flukes}.  We include it  to
see when and why the ACCMM model should agree with free quark decay.

Heavy quark decay proceeds via the Hamiltonian
\beq
H={G_F \over \sqrt{2}} V_{ib} J_i^\mu J_{l\mu}^\dagger
\eeq
where
\begin{eqnarray}
J_i^\mu&=&\overline{q}_i \gamma^\mu(1+\gamma_5) b\\
J_l^\mu&=&\overline{\nu}_l \gamma^\mu(1+\gamma_5)l^-
\end{eqnarray}
where $V_{ib}$ is the KM matrix element .
The dependence on the hadronic
matrix element in the inclusive decay will be of the form
\beq
W_{\mu\nu}=(2\pi)^3\sum_i \delta^4(p_{B}-q-p_X) \langle B |J_\mu^\dagger
|X\rangle \langle B|J_\nu|B\rangle
\eeq

Now for fixed $q^2$, one can study the following quantity
\beq
T_{\mu\nu}(q)=-i\int d^4 x e^{-i q \cdot x} \langle B |T(J_\mu^\dagger(x)
J^\nu(0) |B\rangle
\eeq

The quantity $T_{\mu\nu}$ has the property that the discontinuity across a cut
gives
$W_{\mu\nu}$. That is, ${\rm Im}T_{\mu\nu}=-\pi W_{\mu\nu}$.
There is a cut in $T_{\mu \nu}$ which extends between
$\sqrt{q^2}\le q\cdot v\le(m_b^2+q^2-m_C^2)/(2m_b)$
where  $m_C$ is the mass of the lightest hadron
containing the $c$ quark. There are other cuts corresponding to scattering
processes which are irrelevant to this analysis.

The amplitude $T_{\mu\nu}$ can be  perturbatively computed in a region far
from singularities where the operator product expansion applies.  The matrix
element of the time ordered product of currents
can be expanded in terms of the matrix elements of operators of the heavy quark
theory. The
coefficients of the operator are determined by evaluating the matrix elements
of the
time ordered product between quark and gluon states. The quantity itself is
determined from the matrix elements of the heavy quark operators between the
meson states.

At any given order in $\Lambda_{QCD}/m_b$,  $T_{\mu\nu}$ is expanded in terms
of a finite number of operators.
The important conclusions which arise from this systematic expansion are 1) At
order  $(\Lambda/m)^0$, you reproduce the free quark result, 2) At subleading
order in $(\Lambda/m)$, there are no corrections. This means the
leading corrections to free quark decay    arise only at order $(\Lambda/m)^2$
and
are therefore   small. In fact, in the heavy quark theory, there are only
two possible operators for $B$ mesons. However, one of the operators has an
unknown
matrix element,
so the leading  correction is not known, but has been evaluated in terms
of the unknown parameter (see \cite{shif,mw}).

The key ingredient to the vanishing of the $(\Lambda/m)$ correction was the use
of the equations of motion.  When these were used, there was an implicit
assumption
\cite{fluken} that  the heavy quark field, defined  at leading order in
$(\Lambda/m)$ by
\beq
b(x)=e^{-im_b v\cdot x} b_v(x)
\eeq
was defined in such a way that a potential mass counterterm vanished. That is,
the choice of $m_b$  corresponds to the choice of counterterm $\langle
B|\overline{b_v}\delta m b_v|B \rangle=0$ \cite{fluken}.
There is a single physical parameter, namely the $b$ quark mass. Notice that
for spectator decays, everything depends on this  quark mass (not the meson
mass).
 This means for example
that the rate is the same for the meson or the baryon containing a single $b$
quark.

However, there are limitations to the OPE approach.  The OPE breaks down
at the endpoint of the lepton spectrum. This is because the OPE does
not converge at this point.  This
is apparent when higher order terms are included.
Without smearing, the perturbative expansion in $(\Lambda/m)$ does
not work in the endpoint region. In terms of the dual picture, the endpoint
region
is the regime
which one expects to be resonance dominated. That is,  this is the regime where
the spectrum depends on the details of the bound state.
In order to obtain predictions applicable to this region, a suitable smearing
of energy
must be applied. In either picture, one expects the range of energies which
need to be averaged over
to be of order $\Lambda_{QCD}$.  In the next section, we  explore these
statements in the context of the ACCMM model.

\section{The ACCMM Spectrum}

In this section, we consider the spectrum of the ACCMM model in its entirety.
We compare this model to the statements in the previous section. In particular,
we will be interested to see when averaging is required, and if it is, how
large a range of energy should be averaged to get good agreement?
Of course the important question is really  how well should we expect  the
predictions
of different models to agree, given that the $b$ quark is not really free.

{}From the analysis of the previous section, we expect that for a sufficiently
well averaged spectrum, we will get  agreement with a free quark model
(up to order $(\Lambda/m)^2$), but that the detailed forms of the spectrum
near the endpoint will differ. Both these statements are true, as can be
seen in Figures ~1 and 2.
In this section, we first consider the total integrated spectrum. We then
discuss in detail how the ACCMM model differs from the free quark prediction.
We  averaged over energy to see how large a smearing  is required
to get good agreement (of course
this depends on the accuracy desired). Finally, we discuss the question of how
large
an error in extracting physical parameters one is likely to make by using
one model rather than the other.

First let's consider the total integrated spectrum.
{} From the
analysis of the previous section, we know that the integrated spectrum
should agree well between the full result and our approximation of free $b$
quark decay. This follows because the full integration is certainly averaged
over energy over a sufficiently large interval. Therefore, the
two should agree to within order $(p_f/m_B)^2$. For example, for  $p_f=.3$, the
difference would be expected to be of order $.4\%$. In  Table 1,  for various
values of $p_f$ (with $m_{sp}$ fixed at .15), we
give the corresponding values of $m_b$ and the  lifetime according to the
semileptonic decay of a quark of this mass.
 In
the fourth column, we give
the fractional deviation in percent , $\Delta \tau$ between  the ACCMM model
determined
lifetime
and that of the free quark model.
We see that there is very good agreement.

\begin{table}
\[ \begin{array}{|c|r|r|r|r|}  \hline
 & m_b & X &  \Delta \tau  & \Delta \\ \hline
 p_f=0.1 & 5.02 &  1.733 & 0.1  & 0.5 \\ \hline
 p_f=0.2 & 4.94 &  1.929 & 0.4  & 2.1 \\ \hline
 p_f=0.3 & 4.84 &  2.193 & 0.8  & 4.8 \\ \hline
 p_f=0.4 & 4.73 &  2.522 & 1.2  & 7.4 \\ \hline
 p_f=0.5 & 4.62 &  2.926 & 1.6  & 10.86 \\ \hline
\end{array}
\]
{\figfont  Table 1:
For the values of $p_f$ in column 1, we display the corresponding value of
$m_b$ in the
second column and
the parameter $X$ defined by $\tau =X  \; 10^{-14}\; V_{cb}^{-2} B_{SL}$ (for
a quark of this mass, where $B_{SL}$ is the semileptonic branching fraction) in
the
third column.
In the fourth column is given the fractional discrepancy
in percent between the ACCMM determined lifetime  and  that from free quark
decay,
$\Delta \tau =X  \; 10^{-14}\; V_{cb}^{-2} B_{SL}$. In the last
column is given the deviation  $\Delta = 100 \; \int_0^{E_{max}}
|\frac{d\Gamma_1}{dE}-\frac{d\Gamma_2}{dE}| dE/\Gamma_1$. In all cases we have
taken  $m_{sp}=0.15, \; m_{final} =1.5 $.}
\end{table}

It is easy to see that this had  to be the case.
Since the total spectrum for the
ACCMM model is given by
\begin{eqnarray}\nobreak
\int {d\Gamma_B \over dE} dE&=&\int dp p^2 \phi(|p|) \int {d \cos \theta_p
\over 2} {1 \over \gamma}\times
\nonumber
\\ &&\int dE' {d \Gamma(W,E') \over d E'} \delta(E-\gamma E'-\gamma \beta E'
\cos \theta_p)
\end{eqnarray}
which it is easy to see is the average of $d \Gamma(W,E) /dE$ (up
to quadratic corrections). Again
the average of this quantity agrees with the quantity evaluated at the average
value up to higher order corrections.

However, the entire
spectrum does not agree so well.  In particular,  within about  $2 p_f$ of the
endpoint, the spectra look different.
In
the fifth column is shown the ratio of the integrated absolute value
of the difference of the differential spectra divided by the total integrated
spectrum. We see that this discrepancy is relatively large. That is
because the deviation  can be of order $p_f/m$
(note that it vanishes when $p_f$ goes to zero where the distribution
reduces to a delta function).
It is only because the integrated spectrum averages the regions of positive
and negative difference between the two models that the total
integrated rates are in better agreement.

We gain some insight into this cancellation
by considering the exact expression for the differential decay rate, which is
\begin{eqnarray}
{d \Gamma_B \over dE}&=&\int _0^{p_1}dp p^2 \phi(|p|) {W \over 2p}
\int_{E_-}^{E_+}
{d \Gamma(W,E') \over E'} \\
&+&\int_{p_1}^{p_{max}} dp p^2 \phi(|p|) {W \over 2p} \int_{E_-}^{E_1}
{d \Gamma(W,E') \over E'}
\end{eqnarray}
where $E_1=W(1-\epsilon)/2$ and $p_1=m_B/2(1-\frac{m_f^2}{m_B^2})-E$ for
$m_{sp}=0$.
This spectrum extends up to
$E_{max}=\frac{m_B-m_{sp}}{2}(1-\frac{m_f^2}{(m_B-m_{sp})^2})$.

Notice that near the endpoint,  (this is actually quite a large
region, of order at least $2p_f$), the result differs from the free quark decay
spectrum due to two effects. First is that the term which would have
reproduced
the free quark decay spectrum (the first term) is no longer integrated over
$p$ up to $p_{max}\approx \infty$ but is only integrated to $p_{max}-E$.
For $E$ approaching the endpoint, the range of $p$ integration is smaller
than is necessary to get the full free quark decay spectrum. This reduces
the spectrum of the exact result relative to the free quark decay result
as we approach the endpoint.  However, as $E$ approaches the endpoint,
the second term gains significance. Because the spectrum is rapidly
falling, the expansion in $p$ followed by an average is not   legitimate. Small
values of $p$
are favored, as these are equivalent to effectively larger
quark mass. The integration over effective quark masses,
some of which are larger than $m_b$, permit the decay spectrum to continue
beyond the endpoint of $b$ quark decay. Near the endpoint, the second effect
increases the spectrum over the free quark decay result.

Of course the spectrum near the endpoint is not well predicted, either by QCD
or by models.
It is for this reason that it is very unreliable.
To extract a reliable value of $m_b$,
one could avoid this region and model the remaining measured
parts of the spectrum with free $b$ quark decay. Alternatively, one can average
over energy.  To
see how large an energy interval is required (in this model), we averaged both
the model and the free quark prediction using the averaging function
\beq
{d \widetilde{\Gamma} \over dE}(W,E_0)=\int{1 \over \sqrt{\pi}\Delta E}
e^{-{(E-E_0)^2 \over (\Delta E)^2}}
{d \Gamma(W,E) \over dE} dE
\eeq
Notice that an argument similar to that given above
would show that so long as the averaged spectrum varies
sufficiently slowly, one would expect the average (in momentum) of
the smeared spectrum to be the smeared spectrum evaluted at the average $W$,
so we should expect good agreement for sufficient smearing.

\begin{figure}
\let\picnaturalsize=N
\def\picsize{4.0 in}
\def\picfilename{pf03de06.ps}
\ifx\nopictures Y\else{\ifx\epsfloaded Y\else\input epsf \fi
\let\epsfloaded=Y
\centerline{\ifx\picnaturalsize N\epsfxsize \picsize\fi
\epsfbox{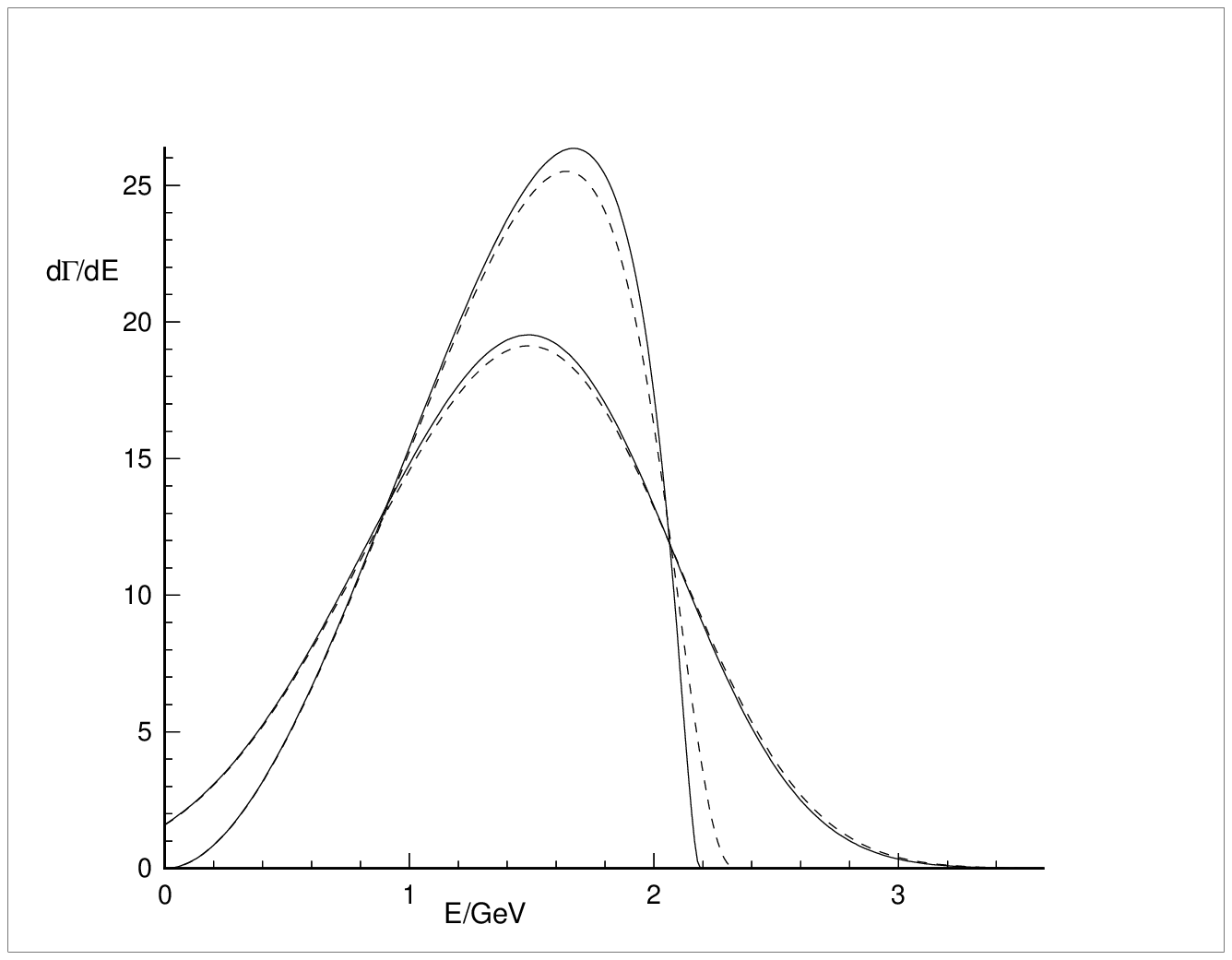}}}\fi
{\figfont  Figure 3: The unaveraged inclusive differential semileptonic decay
rates (in units of $10^{-12}V_{cb}^2$) for  the free quark and ACCMM model for
$p_f=0.3,\; m_{sp}=0.15,\; m_f=1.5$ and the smeared differential decay spectra
for $\Delta E=0.6 {\rm GeV}$. The upper two curves are the unaveraged ones;the
solid curves correspond to the free quark model and the dashed ones to the
ACCMM model.}
\end{figure}

The averaged spectra for $p_f=0.3$ and $\Delta E=0.6$ are displayed in Figure
3.
We see that they agree rather well.
This model yields some insight into how large a smearing
region is necessary.  In order to
the endpoint to get results
accurate at the level of about $(p_f/m_b)^2$, one must smear the result over a
range of energies at least
about $2 p_f$. This can be seen in Table 2 where we give $\Delta$  for  various
values of the smearing parameter $\Delta E$, for $p_f=0.3$. Because $p_f$ is
{\it a priori} an unknown parameter,
the required region is at least of order 500--1000 MeV.

\begin{table}
\[
\begin{array}{|c|r|r|r|r|r|} \hline
&\Delta E=0&\Delta E=0.3&\Delta E=0.6&\Delta E=0.9 &\Delta E=1.2 \\ \hline
\Delta & 4.8&2.7&1.8&1.4&1.1 \\ \hline
\end{array}
\]
{\figfont Table 2: The deviation of the averaged spectra $\Delta$
(defined in Table 1)   for different averaging values
$\Delta E$, in GeV. The fixed parameters are $p_f=0.3,m_{sp}=0.15,m_f=1.5$.}
\end{table}

\begin{table}
\[
\begin{array}{|c|r|r|r|r|r|} \hline
 & 0 & -0.2 & -0.4 & -0.6 & -0.8 \\ \hline
\Delta E=0 & 26 & 20 & 19 & 22 & 29 \\ \hline
\Delta E=0.3 & 7.0 & 3.9 & 4.0 & 7.3 & 13 \\ \hline
\Delta E=0.6 & 2.0 & 0.85 & 1.6 & 4.4 & 9.1 \\ \hline
\Delta E=0.9 & 0.80 & 0.27 & 1.0 & 3.0 & 6.1 \\ \hline
\Delta E=1.2 & 0.38 & 0.12 & 0.69 & 2.0 & 4.2 \\ \hline
\end{array}
\]
{\figfont Table 3: $\chi^2$ for the averaged and unaveraged spectra  for
$p_f=0.3$ where   $ \chi^2 \propto \int_0^{E_{max}} (f_1-f_2)^2 \; f_2 dE$,
where $f_1$ is a free quark spectrum and $f_2$ is the ACCMM spectrum. The
different rows  correspond to different averaging, and the different columns to
the variation of the b quark mass (in percent). The absolute scale is
arbitrary.}
\end{table}

However such precise agreement is perhaps not necessary.
To understand how large an error is made by assuming a free quark model and
neglecting higher order effects, we studied the question of how badly one would
do in extracting  a $b$ quark mass assuming the ACCMM model
were the correct description of a $B$ meson.
This gives some idea about how inaccurate the
extraction of parameters would  be by modeling the meson as
free quark decay.  We defined
$\chi^2=\int_0^{E_{max}} (f_1-f_2)^2 f_2 dE_1$, where
$f_1$ is a free quark spectrum and $f_2$ is the spectrum
from the ACCMM model. We assumed the value of the quark mass one
extracts would correspond to the minimum of this function, and varied
the quark mass by fractional amounts
given in  percent  in Table 3. The extracted
$b$ quark mass would always be a little smaller than the ``correct" one,
because the statistics are dominated by the region away from the endpoint
where the quark model prediction is slightly less than that of the ACCMM model.
Nonetheless,  the error in mass is quite small.
In Table 3,  we have taken $p_f=0.3$ and smeared over
different energies, $\Delta E$. We see that
even with no smearing,
we
would obtain a value of $b$ quark mass which is
accurate at about the $0.5\%$ level.  With smearing,
one can do even better. With a smearing of $\Delta E=.6$,
one might obtain a $b$ quark mass which is correct
at the $0.2 \%$ level. Of course, there are no precise statistics
behind this estimate. Nonetheless, it seems that free quark
decay, even unaveraged, without higher order corrections,
 should describe the $b$ quark decay
spectrum  very well. Whether this is the case will of course depend
on details and the required accuracy.

\section{Conclusion}

A failing of the ACCMM model is that it does not incorporate properly the
parameters of QCD. If one fixes $m_{sp}$ (as was done in ref. \cite{accmm}),
then the only real parameter of the model is $p_f$.
As we have emphasized, this parameter is exchangeable for the $b$ quark mass.
But then there are no free parameters left to characterize Fermi motion and
the spin dependent
operator which occurs at second order. The second operator is not included
at all; the first operator has a coefficient which is determined from $p_f$, or
equivalently
$m_b$.

The relation between $p_f$ and the parameter $K$ of ref \cite{shif,mw} is
\beq
|K|=\langle {p^2 \over 2 m_b^2}\rangle={3\over 4}{p_f^2\over m_b^2}
\eeq
{}From this point of view the values of $K$ used in ref. \cite{shif,mw}
correspond
to fairly large values of $p_f$. Of course the value of $p_f$ is unknown so
these
large values are potentially valid. The values of $p_f$ extracted in ref.
\cite{accmm}
were determined not by the value of Fermi motion, but from the best quark mass
to fit the data. In reality, Fermi motion requires an independent parameter.

However, the ACCMM model does include one additional parameter, namely
$m_{sp}$.
Varying this parameter can change the relation between the $b$ quark mass and
$p_f$.  However, that the model does not incorporate the physics correctly
is manifest at the level of $(\Lambda/m)^2$ corrections. In the ACCMM model,
the parameters $\ev{p_0^2/m^2}$ and $\ev{(\vec{p})^2/m^2}$ occur as independent
parameters,
both giving rise to order $(\Lambda/m)^2$ corrections.
However, in  a gauge invariant formulation, the first matrix element would
correspond to the
expectation value of ${D_0^2}$ which is mass suppressed by the equations
of motion.  Gauge invariance
imposes restrictions which are not
necessarily true in a model.  This
will presumably be a problem of any model which does not incorporate gauge
invariance
and in which $\ev{p_0^2}$ does not vanish.

Nonetheless, we see
from Table 4, that independent of the value of   $m_{sp}$, the free quark decay
model agrees well with ACCMM (in fact the deviation from ACCMM is not strongly
dependent on
$m_{sp}$). This is not surprising, since for small  $m_{sp}$, it is
obviously not important, while for large $m_{sp}$, it looks more like a
nonrelativistic quark model.

\begin{table}
\[
\begin{array}{|c|r|r|r|r|} \hline
 & m_b & X & \Delta \tau & \Delta \\ \hline
m_{sp}=0.15 & 4.84 & 2.193 & 0.82 & 4.84 \\ \hline
m_{sp}=0.3 & 4.75 & 2.462 & 1.21 & 4.34 \\ \hline
m_{sp}=0.5 & 4.60 & 3.040 & 1.67 & 4.36 \\ \hline
\end{array}
\]
{\figfont Table 4: The same as in table 1, for different values of $m_{sp}$.
The fixed parameters are  $p_f=0.3, \; m_f=1.5$.}
\end{table}

In summary, we have shown that the $b$ quark mass can be defined so that
the predictions of the ACCMM model agree with that of a free
quark model  at the level of $(\Lambda/m)^2$, so long as one is sufficiently
far
from the endpoint or a suitable energy average is applied.  Only for results
which depend on the detailed form of the
endpoint region would this approximation be inadequate. However
such results would be unreliable in any case. For the purpose
of LEP experiments, free quark modeling might be sufficient . The
data is presumably  effectively smeared by the fragmentation function of  the
$B$ meson,
even without explicit smearing.

 In our analysis, we have ignored perturbative
QCD corrections and higher order effects. The first are
straightforward to incorporate \cite{accmm,qcd}.  To incorporate
the latter as generally as possible would require leaving the Fermi
motion parameter as independent, and including the known spin dependent
operator.
This would allow one to include higher order effects if greater precision
is necessary.
The error in using a free quark model can presumably be estimated by varying
the amount of smearing and the unknown higher order parameter.

\section{Acknowledgements}

We thank Mike Dugan, Bob Jaffe,  Michael Luke,
Nuria Rius, Steven Selipsky, and Vivek
Sharma for useful conversations.
 While we were concluding this paper, we
saw a preprint by G. Baille (UCLA/93/TEP/47)  who
reaches similar conclusions.

\end{document}